\documentclass[format=acmsmall, review=false, screen=true]{acmart}

\usepackage{booktabs} 
\usepackage{amsmath}
\usepackage{graphicx}
\usepackage{epstopdf}
\usepackage{epsfig}
\usepackage{graphics}
\usepackage{array}
\usepackage{float}
\usepackage{multirow}
\usepackage{wrapfig,lipsum,booktabs}

\usepackage[ruled]{algorithm2e} 

\SetAlFnt{\small}
\SetAlCapFnt{\small}
\SetAlCapNameFnt{\small}
\SetAlCapHSkip{0pt}
\IncMargin{-\parindent}

\acmJournal{CSUR}
\acmVolume{-}
\acmNumber{-}
\acmArticle{-}
\acmYear{2023}
\acmMonth{05}

\providecommand{\tabularnewline}{\\}
\floatstyle{ruled}
\newfloat{algorithm}{tbp}{loa}
\providecommand{\algorithmname}{Algorithm}
\floatname{algorithm}{\protect\algorithmname}

\makeatother

\begin{document}
\bibliographystyle{acmtrans}
\title{Algorithmic Security is Insufficient: A Comprehensive Survey on Implementation
Attacks Haunting Post-Quantum Security}

\author{Alvaro Cintas Canto}
\affiliation{%
  \institution{Marymount University}
  \city{Arlington}
  \state{VA}
  \postcode{22207}
  \country{USA}}
\email{acintas@marymount.edu}

\author{Jasmin Kaur}
\affiliation{%
  \institution{University of South Florida}
  \city{Tampa}
  \state{FL}
  \postcode{33620}
  \country{USA}}
\email{jasmink1@usf.edu}

\author{Mehran Mozaffari Kermani}
\affiliation{%
  \institution{University of South Florida}
  \city{Tampa}
  \state{FL}
  \postcode{33620}
  \country{USA}}
\email{mehran2@usf.edu}
\author{Reza Azarderakhsh}
\affiliation{%
  \institution{Florida Atlantic University}
  \city{Boca Raton}
  \state{FL}
  \postcode{33431}
  \country{USA}}
\email{razarderakhsh@fau.edu}

\begin{abstract}
This survey is on forward-looking, emerging security concerns in post-quantum era, i.e., the implementation attacks for 2022 winners of NIST post-quantum cryptography (PQC) competition
and thus the visions, insights, and discussions can be used as a step
forward towards scrutinizing the new standards for applications ranging
from Metaverse/Web 3.0 to deeply-embedded systems. The rapid advances in quantum computing have brought immense opportunities
for scientific discovery and technological progress; however, it poses
a major risk to today's security since advanced quantum computers
are believed to break all traditional public-key cryptographic algorithms.
This has led to active research on PQC
algorithms that are believed to be secure against classical and powerful
quantum computers. However, algorithmic security is unfortunately
insufficient, and many cryptographic algorithms are vulnerable to
side-channel attacks (SCA), where an attacker passively or actively
gets side-channel data to compromise the security properties that
are assumed to be safe theoretically. In this survey, we explore such
imminent threats and their countermeasures with respect to PQC. We
provide the respective, latest advancements in PQC research, as well
as assessments and providing visions on the different types of SCAs.

\end{abstract}


 \maketitle

\section{Introduction}
Shor's algorithm is a known quantum algorithm that allows solving
discrete-logarithm and integer-factorization problems, making public
key cryptographic standards vulnerable under the presence of quantum
computers. RSA, DSA, and elliptic
curve cryptography (ECC) are the main public key cryptographic algorithms
that are used currently. ECC has replaced RSA in many applications
due to its efficient realizations with the same security level. Nevertheless, the introduction of high-performance quantum
computers has increased the need for the creation of public key cryptosystems
which are resistant to the cyber-attacks enabled by quantum-based
computing systems. The National Institute of Standards and Technology
(NIST) announced in late 2016 the commencement of a project to standardize
one or more quantum computer-resistant public-key cryptography and
digital signature algorithms {[}1{]}. After more than five years
and multiple rounds of reviews, NIST has recently, in 2022, chosen
four candidate algorithms for standardization and left four others
for another round of evaluation. Table 1 shows the current state of
the NIST post-quantum cryptography (PQC) standardization process,
where PKE and KEM stand for public-key encryption and key encapsulation
mechanism, respectively. KEMs, unlike general-purpose PKEs, are not
intended for encrypting application data. Instead, they are specifically
created to establish a shared secret between communication partners
in cryptographic protocols such as Transport Layer Security (TLS),
just like the Diffie-Hellman Key-Exchange method, which is currently
one of the best available options.

PQC cryptography englobes five major types: Lattice-based,
code-based, multivariate-based, hash-based, and isogeny-based cryptography.
Lattice-based cryptography mathematical problem is related to lattices,
which are geometric structures formed by repeating patterns of points
in space; code-based cryptography is formed on error-correcting codes,
a technique used to detect and correct errors in data transmission;
multivariate-based cryptography relies on the hardness of solving
equations with multiple variables (there are not multivariate-based
standards or finalists); hash-based cryptography relies on hash functions,
which are one-way functions where any-size input is mapped into a
fixed value; and lastly, isogeny-based cryptography uses isogenies,
which are mappings between elliptic curves. 

Most of the aforementioned PQC algorithms have large designs and complex
operations. This aspect as well as the continuous advancements in
very-large-scale integration (VLSI) technologies make post-quantum
cryptosystems vulnerable to implementation attacks which are commonly
referred to as side-channel attacks (SCAs). SCAs can be divided into
passive or active attacks. Passive attacks are those whose aim is
to exfiltrate sensitive information by analyzing various physical
parameters of the system, e.g., power consumption, timing information,
or electromagnetic radiation. Active attacks, on the other hand, intend
to reveal the internal states of cryptographic implementations by
injecting transient faults into the system, e.g., differential fault
analysis (DFA). The consequences of these attacks range from the exploitation
of sensitive data by third parties to causing an entire system to
malfunction. 

Therefore, it is extremely important and necessary to explore countermeasures against SCAs for securing emerging post-quantum cryptosystems. This survey is on forward-looking, emerging security concerns in post-quantum era, i.e., the implementation attacks for 2022 winners of NIST PQC competition and thus the visions, insights, and discussions can be used as a step forward towards scrutinizing the new standards for applications ranging from Metaverse/Web 3.0 to deeply-embedded systems.

The remainder of this survey is outlined as follows: Section 2 gives
a technical background of the different PQC standards and finalists
that are currently under a fourth evaluation round in the NIST PQC
standardization process; Section 3 comprehensively reviews and analyzes
different types of SCAs that have been implemented as well several
countermeasures to counter them; and lastly, Section 4 concludes
the survey.
\section{Preliminaries}
As shown in Table 1, three out of four standards are lattice-based,
i.e., CRYSTALS-Kyber, CRYSTALS-Dilithium, and FALCON, while one of
them, SPHINCS$^{+}$ is hash-based. The finalists are mostly code-based,
except for SIKE, which is isogeny-based. However, after a classic
attacks on only one core mounted by excellent research works, SIKE's
team acknowledged that SIKE and SIDH are insecure and should not be
used.

\textbf{CRYSTALS-Kyber:} It is the only PQC PKE/KEM that has been
standardized. Its security depends on the hardness of solving the
learning-with-errors (LWE) problem over module lattices.\begin{wraptable}{r}{6.8cm}
\centering{}\caption{Current state of the NIST PQC Standardization Process }{\scriptsize
\begin{tabular}{|c|c|c|c|}
\hline 
\multirow{2}{*}{PQC Algorithm} & \multirow{2}{*}{Status} & \multirow{2}{*}{Type} & PKE/KEM \emph{vs.}\tabularnewline
 &  &  & Signature\tabularnewline
\hline 
\hline 
CRYSTALS-Kyber  & \multirow{4}{*}{Standard} & \multirow{3}{*}{Lattice} & PEK/KEM\tabularnewline
\cline{1-1} \cline{4-4} 
CRYSTALS-Dilithium &  &  & \multirow{3}{*}{Signature}\tabularnewline
\cline{1-1} 
FALCON &  &  & \tabularnewline
\cline{1-1} \cline{3-3} 
SPHINCS$^{+}$ &  & Hash & \tabularnewline
\hline 
\hline 
BIKE & \multirow{3}{*}{Round 4} & \multirow{3}{*}{Code} & \multirow{4}{*}{PKE/KEM}\tabularnewline
\cline{1-1} 
Classic McEliece &  &  & \tabularnewline
\cline{1-1} 
HQC &  &  & \tabularnewline
\cline{1-3} \cline{2-3} \cline{3-3} 
SIKE & Broken & Isogeny & \tabularnewline
\hline 
\end{tabular}}
\end{wraptable} Both PKE
and KEM are very similar; however, the KEM uses a slightly tweaked
Fujisaki\textendash Okamoto (FO) transform. The LWE problem involves
finding a small secret vector $s$ (secret key) when given a matrix
$A$ over a constant-size polynomial ring and a vector $b=As+e$.
To encode a message $m$, a particular seed value $\mu$ is used,
binomial sampling is employed to select random values ($r$, $e_{1}$,
$e_{2}$), and a uniform distribution is used to sample $A^{T}$.
The values of $u$ and $v$ are then calculated by combining these
elements with the message. Lastly, the ciphertext $c$ is formed by
compressing $u$ and $v$ using a compression algorithm. To decode
the message, an approximation of $v$ is recovered by computing the
product of the secret key and $u$. CRYSTALS-Kyber requires polynomial
ring multiplications and it uses number-theoretic transform (NTT),
which is an efficient way to perform multiplications in lattice-based
cryptosystems; however, it is one of the major vulnerable points against
SCA. 

\textbf{CRYSTALS-Dilithium: }Its security is based on the hardness
of finding short vectors in lattices, known as the Shortest Vector
Problem (SVP), and it operates over the ring $Z_{q}[X]/(X^{n}+1)$
with $q=2^{23}-2^{13}+1$ and $n=256$. The key generation algorithm
creates a matrix $A$ and secret key vectors $s_{1}$ and $s_{2}$
in such polynomial ring. The public key is then computed as $t=A\cdot s_{1}+S_{2}$.
The bulk of the signing and verification procedures in CRYSTALS-Dilithium
involve two operations: expanding an XOF (eXtendable Output Function)
using SHAKE-128 or SHAKE-256, and performing polynomial ring multiplication
using NTT. To compute the signature, a masking vector of polynomials
$y$ is multiplied with $A$ (where $w_{1}$ are the high-order bits),
and a challenge $c$ is then created as the hash of the message and
$w_{1}$. Lastly, the potential signature is computed as $z=y+c\cdot s_{1}$.
Then, the verifier uses the public key and computes $w_{1}^{'}$ to
be the high-order bits of $A\cdot z-c\cdot t$ and accepts the signature
if all coefficients of $z$ are less than a threshold.

\textbf{FALCON:} One of the major drawbacks of CRYSTALS-Dilithium
is their large size signatures. Therefore, FALCON, another lattice-based
cryptosystem whose signatures are of smaller size, has been standardized.
Its underlying hard problem is the short integer solution problem
(SIS) over NTRU lattices. The FALCON scheme is based on a GPV framework,
which provides a way to construct signature schemes using lattice-based
primitives. Another important aspect of FALCON is the use of Fast
Fourier sampling, which improves FALCON's efficiency and performance.
In the key generation, two random polynomials $f$ and $g$ are chosen
and the NTRU equation is solved to find a matching $F$ and $G$.
To sign, the message is hashed along with a random nonce, into a polynomial
$c$ modulo $\phi$, where $\phi=x^{n}+1$ ($n$ is typically 512
or 1024). The signer then uses the secret lattice basis ($f$, $g$,
$F$, $G$) to produce a pair of short polynomials ($s_{1}$, $s_{2}$)
such that $s1=c-s_{2}h\text{ mod }\phi\text{ mod }q$ (where $h=g/f$
and $q=12289$) and $s_{2}$ is the signature. The verifier needs
to recompute $s_{1}$ from $c$ and $s_{2}$ and verify that $(s_{1},s_{2})$
is an appropriately short vector.

\textbf{SPHINCS$^{+}$: }It is the only stateless hash-based PQC cryptosystem
that has been standardized to avoid relying only on the security of
lattices for signatures. Depending on the hash function that SPHINCS$^{+}$
is instantiated with, there are three different schemes: SPHINCS$^{+}$-SHAKE256,
SPHINCS$^{+}$-SHA-256, and SPHINCS$^{+}$-Haraka. SPHINCS$^{+}$
is constructed using a Merkle tree structure where its leaves are
the hash values of the message to be signed. First, the public and
private keys are generated using a deterministic algorithm that takes
a seed of length $n$ as input. SPHINCS$^{+}$ iteratively hashes
and concatenates the leaf values with the intermediate values of the
Merkle tree until the root value is obtained. The root value is then
signed using the private key to generate the signature. 

\textbf{BIKE: }BItFlipping Key Encapsulation, or BIKE, may be regarded
as the utilization of quasi-cyclic moderate density parity check (QC-MDPC)
codes to instantiate the McEliece cryptosystem, using the equivalent
Niederreiter scheme. BIKE has three different variants targeting two
different security properties: Chosen plaintext attacks (CPA) and
chosen ciphertext attacks (CCA) security. The key generation is almost
identical in both the PKE and KEM processes. First, the secret key
$sk$ is formed by two low-weight vectors $h_{0}$ and $h_{1}$ of
length $r$ that are uniformly picked from a secret key space $H_{w}$
(a value $\sigma$ is also used in case there is an error in the decapsulation
KEM process). Then, the public key is computed as $h=h_{1}h_{0}-1$.
For the PKE encryption process, the plaintext is represented by the
sparse vector $(e_{0},e_{1})$, and the ciphertext by its syndrome
$s$ obtained by following $s=e_{0}+e_{1}\cdot h$. To decrypt it,
a Black-Gray-Flip (BGF) decoder, which is defined in {[}2{]}, is used
to obtain the plaintext such as $Decoder(s\cdot h_{0},h_{0},h_{1})$. In the KEM encapsulation process, a random bitstring $m$ is selected
and hashed, obtaining an error vector $(e_{0},e_{1})$ of weight $t$.
A ciphertext $c$ is then calculated in two parts such as $c=(e_{0}+e_{1}\cdot h,m\oplus L(e_{0},e_{1}))$,
where $L$ is a hash function. Lastly, a shared key $K$ is obtained
by hashing $m$ and $c$. In the KEM decapsulation process, the shared
key is obtained by using $c$ and $sk$. First, an error vector $e'$
is obtained by $e'=Decoder((e0+e1h)h0,h0,h1)$.
Then, $m'=(m\oplus L(e_{0},e_{1}))\oplus L(e')$;
 if $e'$ matches $H(m)$ then $K$ is
$K(m',c)$, otherwise $K$ is $K(\sigma,c)$.

\textbf{Classic McEliece: }It is a code-based cryptosystem based on
binary Goppa codes and is widely regarded as secure. However, its
large public key size is not desirable. McEliece generates a pair
of keys using a code subspace dimension $m$, a maximum number of
errors that can be corrected $t$, and a code length $n$. The private
key consists of a monic irreducible polynomial called the Goppa polynomial
with degree $t$, which is generated randomly and all its coefficients
are elements of a finite field $GF(2^{m})$. The public key is obtained
by constructing a control matrix $H$ based on the private key, permutating
it using a random permutation matrix $P$, and transforming it into
a systematic form $G$. To encode a plaintext message $p$, a random
error vector $e$ of length $n$ and weight $t$ is created and the
ciphertext $c$ is calculated as $c=p\cdot G\oplus e$. To decode
the ciphertext, the error vector $e$ is first located using an error
locator polynomial $\sigma(x)$ and then the original plaintext is
reconstructed.

\textbf{HQC: }Hamming Quasi-Cyclic, or HQC, is an efficient encryption
scheme based on coding theory. To have smaller keys than other code-based
cryptosystems, HQC uses two different types of codes: A decodable
$[n,k]$ code $C$ with a fixed generator matrix $G\in F_{2}^{k\times n}$
and error correction capability based on concatenated Reed-Muller
and Reed-Solomon codes, and a random double-circulant $[2n,n]$ code
with a parity check matrix $h$. In the key generation for both PKE
and KEM, the parity check matrix $h$ in $R$ is generated and the
secret key $sk$ is created using polynomials $x$ and $y$ in $R^{2}$.
Next, the public key $pk$ is set such as $pk=(H,s=x+H\cdot y)$.
In the KEM process, three hash functions, named $G$, $K$, and $H$,
are required. To encapsulate any random generated message $m$, the
randomness $\theta$ for the encryption is first derived by $G(m)$.
Then, a ciphertext $c$ is generated by encrypting $m$ using $pk$
and $\theta$. Lastly, a symmetric key $K$ is derived such as $k=K(m,c)$
and the other party receives $(c,d)$, where $d=H(m)$. To decapsulate,
$c$ is decrypted using $sk$, obtaining $m'$.
To verify the integrity of $c$, $m'$
is re-encrypted using $\theta'$, obtain
another ciphertext $c'$. If $c'$
matches $c$ and $d$ matches $H(m')$,
then $K(m,c)$ is the shared key. On the other hand, in the PKE process,
vectors $r_{1}$, $r_{2}$, and $e$, with a fixed hamming weight,
are first sampled. Then, the ciphertext $c$, which is a tuple $e$
with $u=r_{1}+h\cdot r_{2}$ and $v=mG+s\cdot r_{2}+e$, is calculated.
To decrypt $c$ and obtain the original message $m$, the term $v-u\cdot y$
is decoded.

\section{SCA against Post-Quantum Algorithms and Countermeasures}

This section evaluates some of the most up-to-date works on SCAs and
respective countermeasures. Tables 2
and 3 show the different PQC scheme variants depending on their
parameters. 
\begin{table*}[t]
\caption{Parameters of different post-quantum PKE/KEM algorithms }

\centering\scriptsize{%
\begin{tabular}{|c|c|c|c|c|c|>{\centering}m{3.5cm}|}
\hline 
\multirow{2}{*}{Algorithm} & Security & $sk$ size & $pk$ size & $ct$ size & $ss$ size & \multirow{2}{3.5cm}{\centering{}Other Parameters}\tabularnewline
 & level & (bytes) & (bytes) & (bytes) & (bytes) & \tabularnewline
\hline 
\hline 
Kyber-512 & 1 & 1,632 & 800 & 768 & 32 & $n$ = 256; $k$ = 2; $q$ = 3,329\tabularnewline
\hline 
Kyber-768 & 3 & 2,400 & 1184 & 1088 & 32 & $n$ = 256; $k$ = 3; $q$ = 3,329\tabularnewline
\hline 
Kyber-1024 & 5 & 3,168 & 1568 & 1568 & 32 & $n$ = 256; $k$ = 4; $q$ = 3,329\tabularnewline
\hline 
\hline 
BIKE-1 & 1 & 2,244 & 12,323 & 12,579 & 32 & $r$ = 12,323; $w$ = 142; $t$ = 134\tabularnewline
\hline 
BIKE-3 & 3 & 3,346 & 24,659 & 24,915 & 32 & $r$ = 24,659; $w$ = 206; $t$ = 199\tabularnewline
\hline 
BIKE-5 & 5 & 4,640 & 40,973 & 41,229 & 32 & $r$ = 40,973; $w$ = 274; $t$ = 264\tabularnewline
\hline 
\hline 
mceliece348864 & 1 & 6,492 & 261,120 & 96 & 32 & $m$ = 12; $n$ = 3,488; $t$ = 64\tabularnewline
\hline 
mceliece460896 & 3 & 13,608 & 524,160 & 156 & 32 & $m$ = 13; $n$ = 4,608; $t$ = 96\tabularnewline
\hline 
mceliece6688128 & 5 & 13,932 & 1,044,992 & 208 & 32 & $m$ = 13; $n$ = 6,688; $t$ = 128\tabularnewline
\hline 
mceliece6960119 & 5 & 13,948 & 1,047,319 & 194 & 32 & $m$ = 13; $n$ = 6,960; $t$ = 119\tabularnewline
\hline 
mceliece8192128 & 5 & 14,120 & 1,357,824 & 208 & 32 & $m$ = 13; $n$ = 8,192; $t$ = 128\tabularnewline
\hline 
\hline 
hqc-128 & 1 & 40 & 2,249 & 4,481 & 64 & $n_{1}$ = 46; $n_{2}$ = 384;

$n$ = 17,669; $w$ = 66 \tabularnewline
\hline 
hqc-192 & 3 & 40 & 4,522 & 9,026 & 64 & $n_{1}$ = 56; $n_{2}$ = 640;

$n$ = 35,851; $w$ = 100\tabularnewline
\hline 
hqc-256 & 5 & 40 & 7,245 & 14,469 & 64 & $n_{1}$ = 90; $n_{2}$ = 640;

$n$ = 57,637; $w$ = 131\tabularnewline
\hline 
\end{tabular}

Parameters for Kyber: $n$ is the polynomial length; $k$ is the size
of polynomial vectors; $q$ is the prime modulus, BIKE: $r$ is the
block size; $w$ is the row weight; $t$ is the error weight, McEliece:
$m$ is the code subspace; $n$ is the code length; $t$ is the guaranteed
error-correction capability, HQC: $n_{1}$ is the Reed-Solomon code
length; $n_{2}$ is the Reed-Muller code length; $n$ is the vectors
dimension; $w$ is the vectors weight.}
\end{table*}
\begin{wraptable}{r}{8.8cm}
\caption{Parameters of different post-quantum signature algorithms.}

\centering\scriptsize{%
\begin{tabular}{|c|c|c|c|c|}
\hline 
\multirow{2}{*}{Algorithm} & Security & $pk$ size & signature & \multirow{2}{*}{Other parameters}\tabularnewline
 & level & (bytes) & size (bytes) & \tabularnewline
\hline 
\hline 
Dilithium2 & 2 & 1,312 & 2,420 & $n$ = 256; $q$ = 8,380,417\tabularnewline
\hline 
Dilithium3 & 3 & 1,952 & 3,293 & $n$ = 256; $q$ = 8,380,417\tabularnewline
\hline 
Dilithium5 & 5 & 2,592 & 4,595 & $n$ = 256; $q$ = 8,380,417\tabularnewline
\hline 
\hline 
FALCON-512 I & 1 & 897 & 666 & $n$ = 512; $q$ = 12,289\tabularnewline
\hline 
FALCON-1024 V & 5 & 1,793 & 1,280 & $n$ = 1,024; $q$ = 12,289\tabularnewline
\hline 
\hline 
SPHINCS$^{+}$-128s & 1 & 32 & 7,856 & $n$ = 16; $h$ = 63; $d$ = 7 \tabularnewline
\hline 
SPHINCS$^{+}$-128f & 1 & 32 & 17,088 & $n$ = 16; $h$ = 66; $d$ = 22\tabularnewline
\hline 
SPHINCS$^{+}$-192s & 3 & 48 & 16,224 & $n$ = 24; $h$ = 63; $d$ = 7\tabularnewline
\hline 
SPHINCS$^{+}$-192f & 3 & 48 & 35,664 & $n$ = 24; $h$ = 66; $d$ = 22\tabularnewline
\hline 
SPHINCS$^{+}$-256s & 5 & 64 & 29,792 & $n$ = 32; $h$ = 64; $d$ = 8\tabularnewline
\hline 
SPHINCS$^{+}$-256f & 5 & 64 & 49,856 & $n$ = 32; $h$ = 68; $d$ = 17\tabularnewline
\hline 
\end{tabular}

Parameters for Dilithium: $n$ is the ring degree; $q$ is the prime
modulus, FALCON: $n$ is the ring degree; $q$ is the prime modulus,
McEliece:$n$ is the size of the hash output and the WOTS$^{+}$ and
FORS signatures; $h$ is the height of each Merkle tree (determines
the number of WOTS$^{+}$ signatures per layer); $d$ is the depth
of the hypertree.}
\end{wraptable}

As previously mentioned, there are two types of attacks: Passive attacks
and active attacks, also known as invasive attacks. For both types
of attacks, the adversary needs to have access to the actual device
where the cryptographic implementation is taking place. Once the adversary
has access to the system, they can passively observe and analyze different
leakages or actively influence it and evaluate their effects as shown
in Fig. 1. In the following subsections, we summarize first the most
common types of SCAs, then we discuss the most well-known countermeasures,
and lastly, we present different SCA attacks found in the literature
and several countermeasures against them for each PQC algorithm.

\subsection{Types of SCAs and Countermeasures}

As mentioned previously, there are many types of SCAs and they can
be either active or passive. Most of the current research focuses
on passive differential power analysis (DPA) by analyzing the power
consumption during one or multiple operations and active differential
fault analysis (DFA); however, there are some other attacks that need
to be considered, e.g., deep-learning-based SCAs to analyze patterns
from the information extracted, and also timing/cache/algebraic/electromagnetic
attacks. Profiling attacks entail the attacker possessing prior knowledge
of the cryptosystem's implementation for training and testing before
the attack. Conversely, non-profiling attacks are characterized by
the attacker's lack of knowledge about the cryptosystem's implementation,
making them more challenging to execute than profiling attacks.

Robust and adaptive countermeasures are essential for secure data
communication against SCAs.\begin{wrapfigure}{r}{0.55\textwidth}
\begin{centering}
\includegraphics[scale=0.3]{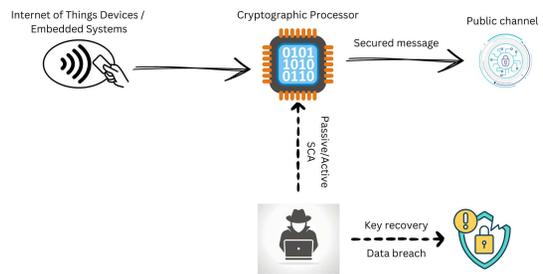}
\par\end{centering}
\caption{SCA representation.}
\end{wrapfigure} These countermeasures can be implemented
as either software-based solutions for passive SCAs or hardware-based
implementations for active SCAs such as fault detection. Passive SCA
countermeasures rely on obfuscating sensitive information via masking
or shuffling to avoid any correlation between the plaintext data and
the information leaked through power consumption, electromagnetic
emissions, or timing variations. The software-based countermeasures
include 1) Algorithmic modifications, such as masking and blinding
techniques, 2) Compiler-based modifications that obfuscate code order,
and 3) Code obfuscation to create incoherence. Hardware-based countermeasures
concentrate on physically securing the cryptographic algorithms against
active SCA by adopting techniques such as power and electromagnetic
shielding (threshold implementation) and error detection/correction
codes to identify fault injections. Another effective strategy against
SCAs involves increasing the system\textquoteright s entropy.

\subsection{CRYSTALS-Kyber SCAs and Countermeasures}

CRYSTALS-Kyber is the only PKE/KEM standardized in the PQC NIST competition
and thus, one of the most evaluated and tested against SCAs. Carrera
\emph{et al.} {[}3{]} propose a non-profiled correlation electromagnetic
analysis against a field programmable gate array (FPGA) implementation
of Kyber-512, recovering the secret subkeys with a success rate of
100\%, given the knowledge of register reference values (not full
knowledge). Ji \emph{et al.} {[}4{]} demonstrate a successful message
(session key) recovery by using a profiling SCA, in particular a deep
learning-based power analysis on a hardware implementation of Kyber768.
All messages with the same enumeration were recovered due to their
novel method called sliced multi-bit error injection. 

Some other attacks target specific building blocks or operations.
In 2017, Primas \emph{et al.} {[}5{]} presented the first single-trace
attack on lattice-based encryption, claiming that a single side-channel
observation is needed for full key recovery. This attack targets the
NTT building block, which is part of the CRYSTALS-Kyber cryptosystem. Xu \emph{et al. }{[}6{]} also targeted the
NTT computation and proposed adaptive electromagnetic SCAs with carefully
constructed ciphertexts, extracting the full secret key with between
8 and 960 traces. Pessl and Primas {[}7{]} changed the target to encryption
to increase the single-trace attack performance. They implemented
a successful attack against CRYSTALS-Kyber on an ARM Cortex M4 microcontroller
assembly-optimized and designed to operate in constant time. Ravi \emph{et
al.} targeted the message decoding by proposing electromagnetic emanation-based
SCAs and fault injection attacks {[}8{]}.

Dubrova \emph{et al.} {[}9{]} perform deep learning-based message
recovery attacks against CRYSTALS-Kyber using a new neural network
training method called recursive learning. To train such neural networks,
in the profiling stage, 30K power traces were collected from the decapsulation
process of different ciphertexts for the same KEM pair and with a
known keypair. The results showed that recovering a message bit from
a single trace of a first-order masked implementation without cyclic
rotations has a probability of $0.127$\%, but with cyclic rotations,
the percentage increases to $87\%$. Furthermore, works {[}10, 11{]}
present side-channel assisted message recovery attacks against CRYSTALS-Kyber
to demonstrate that secret key recovery is possible in shuffled and
masked implementations.

In terms of active attacks (even though some previous works involved
some fault injection), Espitau \emph{et al.} {[}12{]} presented loop-abort
faults on several lattice-based cryptosystems including CRYSTALS-Kyber.
In this attack, a fault is injected into the cryptosystem causing
a loop that samples random Gaussian secret coefficients to abort prematurely.
This premature abortion results in the generation of abnormally low-dimensional
secrets, which can be exploited to carry out a key recovery attack.
However, the actual attack was not carried out for CRYSTALS-Kyber.
In 2021, Pessl and Prokop {[}13{]} presented an attack requiring a
single instruction-skipping fault in the decoding process. Through
fault simulations, they demonstrated that a minimum of 6,500 faulty
decapsulations are necessary to completely recover the key for Kyber512
running on a Cortex M4. Pessl and Prokop claimed that shuffling may
make their attack unsuccessful. Therefore, in the same year, Hermelink
\emph{et al.} {[}14{]} use a combination of fault injections with
chosen-ciphertext attacks against CRYSTALS-Kyber claiming that their
attack may not be mitigated by shuffling the decoder. Their results
show a successful secret key recovery with 7,500 inequalities for
Kyber-512, 10,500 inequalities for Kyber-768, and 11,000 inequalities
for Kyber-1024. A year later, Delvaux {[}15{]} overhauled the SCA
from {[}14{]} to make it easier to perform and harder to protect against
by following four different strategies: Enlargement of the attack
surface; relaxation of the fault model; applying masking and blinding
methods; and accelerating and improving the error tolerance of solving
the system of linear inequalities. 

Several SCA countermeasures have been proposed and a few of them have
been implemented. Masking is one of the most common forms of protecting
CRYSTALS-Kyber against SCAs, especially DPA attacks {[}16, 17, 18,
19{]}. Schneider \emph{et al.} {[}16{]} introduce a secure binomial
sampler that can provide protection against SCAs at any order. This
is achieved through a Boolean and arithmetic (B2A) masking scheme
conversion for prime moduli, suitable for CRYSTALS-Kyber. In {[}17{]}, Bache \emph{et al.} develop a more
efficient higher-order masking scheme for lattice-based schemes with
prime modulus. The scheme is proven in a probing model and tested
on an ARM Cortex-M4F microcontroller, taking only 1.5-2.2 $ms$ to
execute and protecting first-order leakage after collecting 1 million
power traces and applying $t$-test methodology. Bos \emph{et al.}
{[}18{]} also propose a masking implementation for a complete CRYSTALS-Kyber
decapsulation, at both first and higher orders. Their approaches mask
a one-bit compression and decompressed comparison and do not detect
leakage after a Test Vector Leakage Assessment (TVLA) of 100,000 measurements.
Kamucheka \emph{et al.} {[}19{]} also propose a masked pure-hardware
implementation of Kyber-512 and obtain 1.08x and 1.06x overheads in
clock cycles and hardware resources when hiding and masking techniques
are applied. 

Howe \emph{et al.} {[}20{]} propose countermeasures against SCAs that
use the statistical characteristics of the error samples, which are
either Gaussian or binomial. The proposed countermeasures involve
conducting statistical tests to ensure that the samplers are functioning
correctly and take around $85$\% of the overall area consumption.
Ausmita \emph{et al.} {[}21{]} introduce new error detection schemes
based on recomputing and embedded efficiently in the NTT accelerator
architecture on FPGA. The results show a low overhead to detect close
to 100\% of errors. Moreover, also using recomputing, Cintas-Canto
\emph{et al.} {[}22{]} propose error detection schemes for lattice-based
KEMs and implemented them on FPGA. Lastly, Heinz and Poppelmann {[}23{]}
proposed an updated redundant number representation (RNR) approach
to protect CRYSTALS-Kyber's NTT architecture. Furthermore, a novel
DFA countermeasure is derived and implemented using the Chinese Remainder
Theorem (CRT). These techniques aim to protect the arithmetic operations
of lattice-based cryptosystems and obtained a 2.2x computational overhead
when applied to one execution of NTT of the Kyber-768 decryption process.

\subsection{CRYSTALS-Dilithium SCAs and Countermeasures}

As we have seen for CRYSTALS-Kyber, the NTT architecture is a point
of vulnerability against SCAs, especially DPA. While some works explore
attacks on the NTT building block of specific cryptosystems, e.g.,
CRYSTALS-Kyber, they might apply to other lattice-based cryptosystems
that use NTT such as CRYSTALS-Dilithium and FALCON. In {[}24{]}, Steffen
\emph{et al.} presented the first power SCAs of CRYSTALS-Dilithium
in reconfigurable hardware which include: Several profiled simple
power analyses on Dilithium-2 and Dilithium-5 targeting the decoding
and first NTT stage; and a correlation power analysis attack on the
polynomial multiplication. The former had a $94.2$\% success probability
to recover the correct coefficient when using single-trace attacks;
successfully recovered the target coefficient with 50,000 profiling
traces when using multi-trace attacks on decoding; and was capable
of full key recovery with 350,000 profiling traces when using multi-trace
attacks on first NTT stage. In regards to the CPA attack, they successfully
recovered secret coefficients with 66,000 traces.

Before such research, other works on DPA against CRYPTALS-Dilithium
were investigated. In {[}25{]}, Ravi \emph{et al.} proposed a power
analysis attack on the polynomial multiplier in CRYPTALS-Dilithium's
signing process, successfully retrieving a part of the secret key.
Next, Karabulut \emph{et al.} {[}26{]} proposed a single-trace SCA
on $\omega$-small polynomial sampling software that reduces the challenge
of polynomial\textquoteright s entropy for CRYSTALS-Dilithium between
39 to 60 bits. The experiment was done using ARM Cortex-M4F. In the
same year, Marzougui \emph{et al.} {[}27{]} proposed an end-to-end
(equivalent) key recovery attack on CRYSTALS-Dilithium based on a
profiling-based power analysis attack combined with machine learning.
The process only runs sections of the signature process and collects
only the relevant power trace snippet to increase the attack efficiency,
recovering the secret key after tracing the unpack polynomial function
for 756,589 signatures.

In 2018, Bruinderink and Pessl {[}28{]} presented a DFA attack on
deterministic lattice signatures, which included CRYSTALS-Dilithium.
By using linear algebra and lattice-basis reduction techniques, they
show that a single random fault in the signing process can lead to
a scenario of nonce-reuse (enabling key recovery) and that $65.2$\%
of CRYSTALS-Dilithium's execution time is susceptible to an unprofiled
attack. A year later, also pointing out the determinism in lattice-based
signatures, Ravi \emph{et al.} {[}29{]} performed skip-addition fault
attacks targeting the signing operation to extract a portion of the
secret key. Additionally, they introduced a novel forgery method,
enabling an attacker to sign any message using only that portion of
the secret key. In {[}29{]}, the authors also present a zero-cost
mitigation strategy based on re-ordering the operations within the
signing procedure to defend CRYSTALS-Dilithium against their attack,
which increases the attack's time and effort complexity by a $2^{20}$.

Other CRYSTALS-Dilithium SCA countermeasures are found in {[}30, 20,
24{]}. Although there are no specific countermeasures for the CRYSTALS-Dilithium
cryptosystem in {[}30{]}, Bindel \emph{et al.} mentioned several countermeasures
such as masking, switching the order of operands, or storing the result
of the addition in a variable different from the operands, applicable
to several lattice-based signature schemes. Howe \emph{et al.} {[}20{]},
as mentioned earlier, propose fault attack countermeasures based on
statistical tests for error samplers, which are designed to introduce
noise and hide computations on secret information. The work of Steffen \emph{et al.} {[}24{]} also presents
different countermeasures based on arithmetic masking and integration
of decoding into the first NTT stage, being able to protect the CRYSTALS-Dilithium
cryptosystem from the attack previously mentioned.

\subsection{FALCON SCAs and Countermeasures}

While several works perform SCAs against the Gaussian sampling algorithms
used in FALCON, there are not too many specific attacks against the
FALCON cryptosystem. In 2019, McCarthy \emph{et al.} {[}31{]} proposed
the first fault attack on the FALCON signature scheme, using a Basis
Extraction by Aborting Recursion or Zeroing (BEARZ) technique. Through
this attack, it is shown that FALCON is vulnerable to fault attacks
on its Gaussian sampler and the output can reveal the private key.
Moreover, three different countermeasures are proposed in {[}31{]}:
Computing the signature twice, running the verification process immediately
after signing, and applying a zero-check scheme, where the sampled
vector is checked that does not go to zero at some point along its
length at the end of the $ffSampler$ algorithm. The latter is proven
to be the more successful against the SCAs carried in their work. 

A year later, Fouque \emph{et al.} {[}32{]} pinpoint a particular
timing leakage in the FALCON implementations, employing algebraic
number theoretic techniques to retrieve the secret key. Such key retrieval
transpires as a result of information exposure regarding the Gram-Schmidt
norm, a crucial component for converting a group of linearly independent
vectors into an orthonormal basis within the FALCON encryption system.
The Gram-Schmidt process inherently reveals certain numerical properties
of the original vectors allowing the full recovery of the secret key
in FALCON-512. 

Karabulut and Aysu {[}33{]} propose an electromagnetic attack on the
FALCON-512 cryptosystem to extract the secret signing keys by targeting
the floating-point multiplications within FALCON\textquoteright s
Fast Fourier Transform. Their extend-and-prune strategy extracts the
sign, mantissa, and exponent variables without false positives; showing
that \textasciitilde 10k measurements are enough to reveal the secret
key. Guerreau \emph{et al.} {[}34{]} improve the attack of {[}33{]}
in 2022 by exploiting the fact that the polynomial coefficients are
integers. This leads to a reduction of the amount of traces needed
(\textasciitilde 5,000 traces) for full key recovery. Additionally,
they propose a practical but computationally expensive power analysis
of FALCON's Gaussian sampling algorithm, applying a parallelepiped-learning
attack and needing \textasciitilde$10^{6}${} traces for full key
recovery in FALCON-512.

Due to such expense, Zhang \emph{et al.} {[}35{]} have developed several
power analysis attacks on FALCON to significantly lower the requirement
of measurements and computation resources from {[}34{]}. For the first
attack, they discovered that the covariance of the samples in the
slice, i.e., filtered signatures, suffices to reveal the secret, needing
220,000 traces instead of $10^{6}$. Moreover, they perform a practical
power analysis targeting the integer Gaussian sampler of FALCON, relying
on the leakage of random sign flip within the integer Gaussian sampling.
This allows practical key recovery of FALCON-512 with 170,000 traces.

In terms of SCAs countermeasures, besides {[}31{]} and {[}34{]}, which
briefly discuss a small modification of the C code to practically
lower the Hamming weight gap, Sarker \emph{et al.} {[}36{]} provide
error detection schemes based on recomputing for FALCON's sampler.
Such schemes can detect close to 100\% of the errors induced in the
Gaussian sampler. 

\subsection{SPHINCS$^{+}$ SCAs and Countermeasures}

SPHINCS$^{+}$ is the third and last PQC signature algorithm that
has been standardized. The majority of SPHINCS$^{+}$ SCAs have been
active attacks, and research has found that SPHINCS$^{+}$ is the
most sensitive to fault attacks {[}37, 38, 40{]}. Castelnovi \emph{et
al.} proposed the first fault attack on the foundation of the SPHINCS$^{+}$
cryptosystem {[}37{]}. This two-phase attack allows the forgery of
any message signature with just one faulty message. The first stage,
known as the faulting phase, involves requesting two signatures for
the same message. During the computation of the second signature,
a fault is induced, causing a one-time signature (OTS) within the
SPHINCS framework to sign a different value than previously. The subsequent
stage, referred to as the grafting phase, demonstrates that the information
from both signatures\textemdash the accurate one and the faulty one\textemdash can
be utilized to uncover portions of the secret key from the OTS that
experienced the fault, resulting in a partial compromise. The attacker
then exploits this weakened OTS as a means of authenticating a distinct
tree from the one it was intended to authenticate. The assailant generates
a tree entirely under their control and employs the compromised OTS
to graft it onto the SPHINCS tree.

In efforts to provide a practical verification of {[}37{]}, Genet
\emph{et al.} propose the first practical fault attack applied on
an Arduino board for SPHINCS in {[}38{]}, showing how a low-cost injection
of a single glitch is sufficient to obtain exploitable faulty signatures.
In the same year, 2018, Amiet \emph{et al.} {[}39{]} presented the
first hardware-based implementation of SPHINCS$^{+}$ and a fault
attack against such hardware implementation. Amiet \emph{et al.} discovered
that a fault occurring in WOTS$^{+}$ subtree computations results
in an altered root node value. This incorrect root node is subsequently
signed with the next WOTS$^{+}$ level, leaking portions of the associated
WOTS$^{+}$ private key. Consequently, such work demonstrates that,
through a glitch attack, gathering private data to forge a signature
can be accomplished in a matter of seconds. Additionally, a countermeasure
based on doubling the entire SPHINCS$^{+}$ coprocessor is proposed
in {[}40{]}, similar to the recomputing approach suggested by {[}9{]}.
Kannwischer \emph{et al.} entirely exclude fault injection attacks
to analyze the DPA vulnerability of XMSS and SPHINCS {[}40{]}, and
show a practical attack on the BLAKE-256-based PRF used within SPHINCS-256. Other works exclusively focus on providing SCA countermeasures for
SPHINCS$^{+}$ cryptosystem {[}41, 42, 43{]}. Mozaffari-Kermani \emph{et
al.} {[}41, 42{]} propose reliable and error detection hash trees
for stateless hash-based signatures suitable to SPHINCS$^{+}$. Their
work presents two different approaches: Recomputing with swapped nodes
(RESN) in the hash-tree constructions and combined signatures, and
recomputing with encoded operands (REEO) for ChaCha, which is a stream
cipher that SPHINCS uses for deriving two hash functions. The schemes
detected close to 100\% transient and permanent faults, adding up
to 14.6\% degradation overhead on application-specific integrated
circuit (ASIC). The issue with these countermeasures is that they
do not cover the entire SPHINCS$^{+}$ signing procedure. With this
in mind, Genet {[}43{]} introduces a fault attack countermeasure based
on caching the intermediate W-OTS$^{+}$. However, this approach is
useful for stateful schemes such as XMSS$^{MT}$ but not for stateless
schemes such as SPHINCS$^{+}$. Therefore, recomputing schemes are
suggested to be used to protect SPHINCS$^{+}$ against fault attacks
{[}43{]}.

\subsection{BIKE SCAs and Countermeasures}

BIKE can be described as the McEliece scheme instantiated with QC-MDPC
codes. In 2016, Guo \emph{et al.} {[}44{]} introduced an attack using
a recognized correlation between error patterns in decoding failures
and the secret key, under the assumption that the scheme operates
in a static key environment needing IND-CCA security. Such attack
is implemented for 80-bit security QC-MDPC scheme, recovering the
key in minutes. Two years later, an error amplification attack, built
on the previous attack, is proposed {[}45{]}. This attack improves
it by using just a single initial error vector, which leads to a decoding
failure. It then adjusts this vector to efficiently produce numerous
additional error vectors that also result in decoding failures. However,
the attacks from {[}44, 45{]} can be avoided by stronger parameters.

A more recent generic power/electromagnetic attack based on the Fujisaki\textendash Okamoto
(FO) transformation and its variants are proposed by Ueno \emph{et
al.} in {[}46{]}. This attack exploits side-channel leakage during
the non-protected pseudorandom function (PRF) execution in the re-encryption
of the KEM decapsulation and can be applied to CRYSTALS-Kyber, HQC,
and BIKE. 

Since none of these attacks considered the non-constant time rejection
sampling routine, which BIKE and HQC use to generate random vectors
with a specific Hamming weight, Guo \emph{et al.} {[}47{]} propose
two novel timing attacks against BIKE and HQC achieving full secret
key recovery. These attacks examine the time discrepancies caused
by rejection sampling, as they could reveal whether the input message
to the deterministic re-encryption process (or a hash function) in
the IND-CCA transformation remains unaltered. Possessing such secret
information is sufficient for retrieving the secret key of BIKE and
HQC schemes. To fix the non-isochronous design of BIKE, Sendrier {[}48{]} replaces
the rejection sampling in the encapsulation and the decapsulation
with an algorithm that has no rejection, generating a non-uniform
distribution of the indices. Additionally, Drucker \emph{et al.} {[}49{]}
propose to use the fixed sampling number (FSN) version of the errors-vector
generation (EVG), with some predetermined value of X. This value does
not change the required uniform distribution property of the generated
errors-vector. 

Chou \emph{et al.} {[}50{]} also propose a constant-time implementation
for QC-MDPC code-based cryptography to counter timing attacks. Nevertheless,
this countermeasure was later identified as susceptible to a DPA in
private syndrome computation {[}51{]}, although the attack was unable
to fully retrieve the correct secret indices. Thus, Sim \emph{et al.}
{[}52{]} enhance existing multiple-trace attacks on timing attack
countermeasures and propose a novel single-trace attack, which allows
to recover secret indices even when using ephemeral keys or DPA countermeasures.

\subsection{McEliece SCAs and Countermeasures}

Timing
attacks are one of the first SCAs carried on the McEliece cryptosystem
{[}53, 54, 55{]}. Strenzke \emph{et al.} {[}53{]} present a timing
attack on the degree of the error locator polynomial, which is executed
successfully against a software implementation of the McEliece cryptosystem.
Therefore, raising its degree artificially is proposed as a countermeasure.
Avanzi \emph{et al.} improve the timing attack from {[}53{]} with
a setup stage that involves profiling the algorithm for all correctable
error weights, followed by an iterative procedure that approximates
the random error vector. Additionally, a ``non-support'' countermeasure is proposed. In {[}54{]},
Shoufan \emph{et al.} propose a timing attack against the Patterson
algorithm in the McEliece cryptosystem. In {[}56{]}, Lahr \emph{et al.} adapt the side-channel attack from
{[}54{]} and perform an electromagnetic attack using a reaction-based
attack combined with a technique that they call iterative chunking.
This method allows them to progressively increase the quantity of
discovered error positions (chunks) within a single (cumulative) query.
Such attack is performed on a microcontroller targeting the matrix-vector
multiplication of the encryption process and recovering the message
from one faulty syndrome and the public key. A practical evaluation
of the attack is performed on FPGA and it is shown that \textasciitilde 560
measurements are sufficient to mount a successful plaintext recovery
attack. Moreover, Strenzke {[}55{]} develops a strategy how to exploit
a vulnerability in the Patterson algorithm, enabling the attacker
to obtain information about the secret permutation via a timing side
channel. 

Not only timing attacks have been studied, but also fault injection
attacks {[}57, 58, 59{]}, power analysis attacks {[}60, 61{]}, and
message-recovery attacks {[}62{]}. In {[}57{]}, Cayrel and Dusart
present a fault injection attack on different variables of the McEliece
schemes and the possible outcomes are discussed; however, no implementation
is performed. A few years later, Cayrel \emph{et al.} {[}58{]} perform
a message-recovery laser fault injection attack targeting the syndrome
decoding problem on the Classic McEliece cryptosystem. Several experiments
are conducted on a 6-core CPU clocked at 2.8 GHz and 32 GB of RAM
desktop computer to validate the success of the attack, which show
the secret message can be retrieved in less than three seconds. Pircher
\emph{et al.} {[}59{]} recently introduced a key-recovery fault injection
attack targeting the Goppa code's error-locator polynomial and the
decryption algorithm's validity checks, thus making a chosen ciphertext
attack feasible. 

When considering power analysis attacks, Molter \emph{et al.} {[}60{]}
introduced a simple SCA on a McEliece cryptoprocessor using power
analysis. This FPGA-based attack exploits an information leak resulting
from the correlation between the error vector weight and the number
of iterations in the extended Euclidean algorithm used in the Patterson
Algorithm (as in {[}54{]}). In a separate study, Guo \emph{et al.}
{[}61{]} formulated an attack algorithm where unique ciphertexts,
corresponding to single-error cases, are submitted to the decryption
oracle. Decoding these ciphertexts involves only a single entry in
an extensive secret permutation, which forms part of the secret key.
By identifying a leak in the additive FFT step, which is used to evaluate
the error locator polynomial, it is possible to determine a single
entry of the secret permutation. Repeating this process for other
entries results in full secret key recovery. The attack employs power
analysis on FPGA and ARM Cortex-M4, alongside a machine-learning-based
classification algorithm to identify the error locator polynomial
from a single trace. The findings show that full key recovery can
be achieved with less than 800 traces. Lastly, in {[}62{]}, Colombier
\emph{et al.} conduct a side-channel attack by analyzing power consumption
during the matrix-vector multiplication phase of the encryption process.

Other SCA countermeasures are discussed in {[}63, 64, 65, 66, 67,
68, 69{]}. The simple power analysis countermeasure proposed in {[}63{]}
is based on avoiding branch statements and data-dependent timing on
the implementation of the McEliece cryptosystem. This countermeasure
is tested on an ARM Cortex-M3, preventing simple power analysis and
timing attacks but increasing the latency by a factor of 3. In {[}64,
65{]}, natural and injected fault detection schemes based on CRC and
cyclic codes are proposed, respectively. These schemes target the
finite field multipliers used in code-based cryptosystems such as
Classic McEliece and are implemented on FPGA detection close to 100\%
faults. Moreover, Cintas-Canto \emph{et al.} present error detection
schemes based on single parity, interleaved parity, CRC, and Hamming
codes for the $GF(2^{m})$ inversion block {[}66, 67{]} and composite
field arithmetic architectures {[}68{]} that the McEliece cryptosystem
employs. Additionally, fault detection schemes based on CRC are proposed
for the different blocks of the McEliece key generator in {[}69{]}.
After being implemented on FPGA, the schemes detected close to 100\%
of faults and added a worst-case area and delay overhead of 49\%.

\subsection{HQC SCAs and Countermeasures}

Some BIKE SCAs are applicable to the HQC cryptosystem since they share
some operational architectures. One example of this is the work of
Guo \emph{et al.} {[}47{]}, which as it was mentioned before, proposes
two novel timing attacks against BIKE and HQC achieving full secret
key recovery. Another timing attack is presented in {[}70{]} by Huang
\emph{et al.}, in which a cache-timing-based distinguisher for implementing
a plaintext-checking (PC) oracle is presented. This PC oracle employs
side-channel information to verify whether a given ciphertext decrypts
to a specific message. Furthermore, a practical attack is presented
on an HQC execution on Intel SGX, necessitating an average of 53,857
traces for complete key recovery. This attack demands significantly
fewer PC oracle calls than Guo \emph{et al.}'s timing attack in {[}47{]}.

Apart from timing attacks, HQC has also been a target of power analysis
attacks. In {[}71{]}, Schamberger \emph{et al.} propose the first
power SCA on the KEM version of HQC. This attack uses a power side-channel
to create an oracle that determines whether the BCH decoder in HQC\textquoteright s
decryption algorithm rectifies an error for a chosen ciphertext. Considering
the decoding algorithm employed in HQC, it is demonstrated how to
craft queries so that the oracle's response enables the extraction
of a significant portion of the secret key. The remaining part of
the key can subsequently be discovered using a linear algebra-based
algorithm. Experiments show that fewer than 10,000 measurements are
enough to successfully execute the attack on the HQC reference implementation
running on an ARM Cortex-M4 microcontroller. Another power analysis
attack is presented in {[}72{]}, where the authors showcase a novel,
proven power SCA that enables a successful power SCA against the updated
round three version of the HQC cryptosystem (a Reed-Muller and Reed-Solomon
version of HQC). This attack reduces the required attack queries of
{[}47{]} by a factor of 12 and eliminates the inherent uncertainty
of their employed timing oracle. The general idea of the attack is
to choose $v$ such that the decoding result depends on $y_{i}^{(0)}$
(where $y$ is the secret key polynomial), revealing its support.
This attack is also implemented on an ARM Cortex-M4 microcontroller. Lastly, Goy \emph{et al.} {[}73{]} introduce a new key recovery side-channel
attack on HQC with chosen ciphertext. This attack exploits the reuse
of a static secret key on a microcontroller, recovering the static
secret key by targeting the Reed-Muller decoding step of the decapsulation,
specifically focusing on the Hadamard transform. The side-channel
information obtained in the function is used to build an Oracle that
distinguishes between several decoding patterns of the Reed-Muller
codes. Moreover, they show how to query the Oracle such that the responses
give full information about the static secret key. Experiment results
indicate that fewer than 20,000 electromagnetic attack traces are
enough to recover the entire static secret key that the decapsulation
uses. As a countermeasure, the authors propose a masking-based structure
against Reed-Muller decoding distinguisher. 

\section{Conclusion}

Due to the imminent threat that quantum computers pose to current
public-key cryptographic algorithms, there has been a extensive research
on PQC. This survey englobes a comprehensive exploration of PQC, highlighting
that PQC, while designed to be secured against classical and quantum
computers, is still vulnerable to SCAs. These attacks, both passive
and active, are a significant risk as they facilitate key recovery.
This review further elaborates on several forms of SCAs and countermeasures
to mitigate them. It is evident that while advancements in PQC are
significant, the reliability of these algorithms is greatly influenced
by their vulnerability to SCA. Thus, the field of PQC needs ongoing
research and development to ensure not just security from quantum
computing threats, but also reliability against SCAs.

\section*{Acknowledgements}

This work was supported by the US National Science Foundation (NSF) award SaTC-1801488.


\begin{thebibliography}{10}
\bibitem{key-9}D. Moody. Post-Quantum Cryptography:
NIST\textquoteright s Plan for the Future. Feb.
2016.

\bibitem{key-1}N. Drucker, S. Gueron, D. Kostic. QC-MDPC decoders
with several shades of gray. PQCrypto, pp. 35-50, 2020.

\bibitem{key-1}R. C. Rodriguez, F. Bruguier, E. Valea, and P. Benoit. Correlation electromagnetic analysis on an FPGA implementation
of CRYSTALS-Kyber.  Cryptology ePrint Archive, 2022.

\bibitem{key-2}Y. Ji, R. Wang, K. Ngo, and E. Dubrova. A side-channel
attack on a HW implementation of Kyber. ETS, 2023.

\bibitem{key-3}R. Primas, P. Pessl, and S. Mangard. Single-trace
side-channel attacks on masked lattice-based encryption. CHES 2017, pp. 513-533, Springer, 2017.

\bibitem{key-4}Z. Xu, O. Pemberton, S. Roy, D. Oswald, and Z. Zheng. Magnifying side-channel leakage of lattice-based
cryptosystems with chosen ciphertexts: The case study of kyber. 
IEEE Trans. Comput., vol. 71, no. 9, pp. 2163-2176, 2021.

\bibitem{key-5}P. Pessl and R. Primas. More practical single-trace
attacks on the number theoretic transform. LATINCRYPT. vol. 6, pp. 130-149, Springer, 2019.

\bibitem{key-6}P. Ravi, S. Bhasin, S. S. Roy, and A. Chattopadhyay. Drop by drop you break the rock-exploiting generic vulnerabilities
in lattice-based PKE/KEMs using EM-based physical attacks.  IACR
Cryptology ePrint Archives, Report 549, 2020.

\bibitem{key-7}E. Dubrova, K. Ngo, and J. Grtner. Breaking a
fifth-order masked implementation of CRYSTALS-Kyber by copy-paste. 
Cryptology ePrint Archive, 2022.

\bibitem{key-8}L. Backlund, K. Ngo, J. Grtner, and E. Dubrova. Secret
key recovery attacks on masked and shuffled implementations of CRYSTALS-Kyber
and Saber.  Cryptology ePrint Archive, 2022.

\bibitem{key-9}P. Ravi, S. Bhasin, S. S. Roy, and A. Chattopadhyay. On exploiting message leakage in (few) NIST PQC candidates for
practical message recovery attacks.  IEEE Trans. Information
Forensics and Security, vol. 17, pp. 684-699, 2021.

\bibitem{key-10}T. Espitau, P. A. Fouque, B. Gerard, and M. Tibouchi. Loop-abort faults on lattice-based signature schemes and key exchange
protocols.  IEEE Transactions on Computers, vol. 67, no. 11, pp.1535-1549,
2018.

\bibitem{key-11}P. Pessl and L. Prokop. Fault attacks on CCA-secure
lattice KEMs.  IACR Transactions on Cryptographic Hardware and Embedded
Systems, pp. 37-60, 2021.

\bibitem{key-12}J. Hermelink, P. Pessl, and T. Poppelmann. Fault-enabled
chosen-ciphertext attacks on Kyber. INDOCRYPT,
pp. 311-334. Springer, 2021.

\bibitem{key-13}J. Delvaux. Roulette: A diverse family of feasible
fault attacks on masked Kyber.  Cryptology ePrint Archive, 2021.

\bibitem{key-14}T. Schneider, C. Paglialonga, T. Oder, and T. Guneysu. Efficiently masking binomial sampling at arbitrary orders for lattice-based
crypto.  PKC 2019. pp. 534-564, Springer, 2019.

\bibitem{key-15}F. Bache, C. Paglialonga, T. Oder, T. Schneider,
and T. Guneysu. High-speed masking for polynomial comparison in
lattice-based KEMs.  IACR Transactions on Cryptographic Hardware
and Embedded Systems, pp. 483-507, 2020.

\bibitem{key-16}J. W. Bos, M. Gourjon, J. Renes, T. Schneider, and
C. V. Vredendaal. Masking Kyber: First-and higher-order implementations. 
IACR Transactions on Cryptographic Hardware and Embedded Systems,
pp. 173-214, 2021.

\bibitem{key-17}T. Kamucheka, A. Nelson, D. Andrews, and M. Huang. A masked pure-hardware implementation of Kyber cryptographic algorithm. ICFPT. pp. 1-9, 2022.

\bibitem{key-18}J. Howe, A. Khalid, M. Martinoli, F. Regazzoni, and
E. Oswald. Fault attack countermeasures for error samplers in lattice-based
cryptography. ISCAS. pp. 1-5, 2019.

\bibitem{key-19}A. Sarker, A. Cintas-Canto, M. Mozaffari-Kermani,
and R. Azarderakhsh. Error detection architectures for hardware/software
co-design approaches of number theoretic transform.  IEEE Transactions
on Computer-Aided Design Integrated Circuits Systems, accepted, to
appear 2023.

\bibitem{key-20}A. Cintas-Canto, A. Sarker, J. Kaur, M. Mozaffari-Kermani,
and R. Azarderakhsh. Error detection schemes assessed on FPGA for
multipliers in lattice-based key encapsulation mechanisms in post-quantum
cryptography.  IEEE Transactions on Emerging Topics in Computing,
accepted, to appear 2023.

\bibitem{key-21}D. Heinz and T. Poppelmann. Combined fault and
DPA protection for lattice-based cryptography.  IEEE Transactions
on Computers, 2022.

\bibitem{key-22}H. Steffen, G. Land, L. Kogelheide, and T. Guneysu. Breaking and protecting the crystal: Side-channel analysis of Dilithium
in hardware.  Cryptology ePrint Archive, 2022.

\bibitem{key-23}P. Ravi, M. P. Jhanwar, J. Howe, A. Chattopadhyay,
and S. Bhasin. Side-channel assisted existential forgery attack
on Dilithium-a NIST PQC candidate.  Cryptology ePrint Archive, 2018.

\bibitem{key-25}E. Karabulut, E. Alkim, and A. Aysu. Single-trace
side-channel attacks on w-small polynomial sampling: with
applications to NTRU, NTRU prime, and CRYSTALS-Dilithium. HOST, pp. 35-45, 2021.

\bibitem{key-24}S. Marzougui, V. Ulitzsch, M. Tibouchi, and J. P.
Seifert. Profiling side-channel attacks on Dilithium: A small bit-fiddling
leak breaks it all.  Cryptology ePrint Archive, 2022.

\bibitem{key-26}L. G. Bruinderink and P. Pessl. Differential fault
attacks on deterministic lattice signatures.  IACR Transactions
on Cryptographic Hardware and Embedded Systems, pp. 21-43, 2018.

\bibitem{key-27}P. Ravi, M. P. Jhanwar, J. Howe, A. Chattopadhyay,
and S. Bhasin. Exploiting determinism in lattice-based signatures:
Practical fault attacks on pqm4 implementations of NIST candidates. ACM Asia CCS, pp. 427-440, 2019.

\bibitem{key-28}N. Bindel, J. Krmer, and J. Schreiber. Hampering
fault attacks against lattice-based signature schemes: countermeasures
and their efficiency (special session). Hardware/Software Codesign
and System Synthesis Companion, pp. 1-3, 2017.

\bibitem{key-31}S. McCarthy, J. Howe, N. Smyth, S. Brannigan, and
M. O\textquoteright Neill. BEARZ attack FALCON: implementation
attacks with countermeasures on the FALCON signature scheme.  Cryptology
ePrint Archiv, 2019.

\bibitem{key-32}P. A. Fouque, P. Kirchner, M. Tibouchi, A. Wallet,
and Y. Yu. Key recovery from Gram\textendash Schmidt norm leakage
in hash-and-sign signatures over NTRU lattices.  EUROCRYPT, pp. 34-63, Springer, 2020.

\bibitem{key-33}E. Karabulut and A. Aysu. FALCON down: Breaking
FALCON post-quantum signature scheme through side-channel attacks. 
DAC, pp. 691-696,
2021.

\bibitem{key-34}M. Guerreau, A. Martinelli, T. Ricosset, and M. Rossi. The hidden parallelepiped is back again: Power analysis attacks
on FALCON.  IACR Transactions on Cryptographic Hardware and Embedded
Systems, pp. 141-164, 2022.

\bibitem{key-35}S. Zhang, X. Lin, Y. Yu, and W. Wang. Improved
Power Analysis Attacks on FALCON. EUROCRYPT. pp. 565-595, Springer,
2023.

\bibitem{key-36}A. Sarker, M. Mozaffari-Kermani, and R. Azarderakhsh. Efficient error detection architectures for post-quantum signature
FALCON's Sampler and KEM Saber.  IEEE Trans. VLSI Systems, vol. 30, no. 6, pp.
794-802, 2022.

\bibitem{key-37}L. Castelnovi, A. Martinelli, and T. Prest. Grafting
trees: a fault attack against the SPHINCS framework. PQCrypto, Proceedings
9, pp. 165-184, Springer, 2018.

\bibitem{key-39}A. Gent, M. J. Kannwischer, H. Pelletier, and A.
McLauchlan. Practical fault injection attacks on SPHINCS.  Cryptology
ePrint Archive, 2018.

\bibitem{key-40}D. Amiet, L. Leuenberger, A. Curiger, and P. Zbinden. FPGA-based SPHINCS$^{+}$ implementations: Mind the glitch. DSD,
pp. 229-237, 2020.

\bibitem{key-86}M. J. Kannwischer, A. Gent, D. Butin, J. Krmer,
and J. Buchmann. Differential power analysis of XMSS and SPHINCS. COSADE, pp. 168-188, Springer, 2018.

\bibitem{key-42}M. Mozaffari-Kermani, R. Azarderakhsh, and A. Aghaie. Fault detection architectures for post-quantum cryptographic stateless
hash-based secure signatures benchmarked on ASIC.  ACM Transactions
on Embedded Computing Systems, vol. 16, no. 2, pp. 59:1-19,
2016.

\bibitem{key-43}M. Mozaffari-Kermani and R. Azarderakhsh. Reliable
hash trees for post-quantum stateless cryptographic hash-based signatures. DFTS, pp. 103-108, 2015.

\bibitem{key-41}A. Gent. On protecting SPHINCS$^{+}$ against
fault attacks.  Cryptology ePrint Archive, 2023.

\bibitem{key-46}Q. Guo, T Johansson, and P. Stankovski. A key
recovery attack on MDPC with CCA security using decoding errors. ASIACRYPT, pp. 789-815, Springer, 2016.

\bibitem{key-47}A. Nilsson, T. Johansson, and P. S. Wagner. Error
amplification in code-based cryptography.  Cryptology ePrint,
2018.

\bibitem{key-48}R. Ueno, K. Xagawa, Y. Tanaka, A. Ito, J. Takahashi,
and N. Homm. Curse of re-encryption: A generic power/em analysis
on post-quantum kems.  IACR Trans. Cryptographic Hardware
and Embedded Systems, pp. 296-322, 2022.

\bibitem{key-49}Q. Guo, C. Hlauschek, T. Johansson, N. Lahr, A. Nilsson,
and R. L. Schrder. Don\textquoteright t reject this: Key-recovery
timing attacks due to rejection-sampling in HQC and BIKE.  IACR
Trans. CHES, pp. 223-263,
2022.

\bibitem{key-50}N. Sendrier. Secure sampling of
constant-weight words\textendash application to bike.
Cryptology ePrint Archive, 2021.

\bibitem{key-51}N. Drucker, S. Gueron, and D. Kostic. To reject
or not reject: That is the question. The case of BIKE post quantum
KEM. Information
Technology-New Generations, pp. 125-131, Springer, 2012.

\bibitem{key-52}T. Chou. QcBits: constant-time
small-key code-based cryptography. CHES, pp. 280-300, Springer, 2016.

\bibitem{key-53}M. Rossi, M. Hamburg, M. Hutter, and M. E. Marson. A side-channel assisted cryptanalytic attack against
QcBits. CHES,
pp. 3-23, Springer, 2017.

\bibitem{key-54}B.Y. Sim, J. Kwon, K. Y. Choi, J. Cho, A. Park, and
D.-G. Han. Novel side-channel attacks on quasi-cyclic
code-based cryptography. IACR Transactions on
Cryptographic Hardware and Embedded Systems, pp. 180-212, 2019.

\bibitem{key-55}F. Strenzke, E. Tews, H. G. Molter, R. Overbeck,
and A. Shoufan. Side channels in the McEliece PKC. PQCrypto,
pp. 216-229, Springer, 2008.

\bibitem{key-56}A. Shoufan, F. Strenzke, H. G. Molter, and M. Stttinger. A timing attack against Patterson algorithm in the
McEliece PKC. ICISC, pp.
161-175, Springer, 2010.

\bibitem{key-57}F. Strenzke. A timing attack against the secret
permutation in the McEliece PKC.PQCrypto, pp. 95-107,
Springer, 2010.

\bibitem{key-58}N. Lahr, R. Niederhagen, R. Petri, and S. Samardjiska. Side channel information set decoding using iterative chunking:
Plaintext recovery from the Classic McEliece
hardware reference implementation. ASIACRYPT, pp.
881-910, 2020.

\bibitem{key-59}P. L. Cayrel and P. Dusart. McEliece/Niederreiter
PKC: Sensitivity to fault injection. Fut. Inf. Tech., pp. 1-6, 2010.

\bibitem{key-60}P. L. Cayrel, B. Colombier, V. F. Dr\u{a}goi, A.
Menu, and L. Bossuet. Message-recovery laser fault injection attack
on the classic McEliece cryptosystem. EUROCRYPT, pp. 438-467, Springer,
2021.

\bibitem{key-61}S. Pircher, J. Geier, J. Danner, D. Mueller-Gritschneder,
and A. Wachter-Zeh. Key-recovery fault injection attack on the
Classic McEliece KEM. Code-Based Cryptography Workshop, pp.
37-61, Springer, 2023.

\bibitem{key-63}H. G. Molter, M. Stttinger, A. Shoufan, and F. Strenzke. A simple power analysis attack on a McEliece cryptoprocessor. 
Journal of Cryptographic Engineering vol. 1, pp. 29-36, 2011.

\bibitem{key-64}Q. Guo, A. Johansson, and T. Johansson. A key-recovery
side-channel attack on classic McEliece. ePrint,
2022.

\bibitem{key-65}B. Colombier, V. F. Dr\u{a}goi, P. L. Cayrel, and
V. Grosso. Profiled side-channel attack on cryptosystems based
on the binary syndrome decoding problem.  IEEE Trans. Information
Forensics and Security, vol. 17, pp.3407-3420, 2022.

\bibitem{key-66}M. Petrvalsky, T. Richmond, M. Drutarovsky, P. L.
Cayrel, and V. Fischer. Countermeasure against the SPA attack on
an embedded McEliece cryptosystem. RADIOELEKTRONIKA, pp. 462-466, 2015.

\bibitem{key-73}A. Cintas-Canto, M. Mozaffari-Kermani, R. Azarderakhsh. Reliable CRC-based error detection constructions for finite field
multipliers with applications in cryptography.  IEEE Trans. VLSI Systems, vol. 29, no. 1, pp.
232-236, 2021.

\bibitem{key-67}A. Cintas-Canto, M. Mozaffari-Kermani, and R. Azarderakhsh. Reliable architectures for finite field multipliers using cyclic
codes on FPGA utilized in classic and post-quantum cryptography. IEEE Trans. VLSI Systems, vol.
1, no. 31, pp. 157-161, 2023.

\bibitem{key-74}A. Cintas-Canto, M. Mozaffari-Kermani, and R. Azarderakhsh. CRC-based error detection constructions for FLT and ITA finite
field inversions over $GF(2^m)$.  IEEE Trans. VLSI Systems, vol. 29, no. 5, pp.
1033-1037, 2021.

\bibitem{key-68}A. Cintas-Canto, M. Mozaffari-Kermani, and R. Azarderakhsh. Error detection constructions for ITA finite field inversions over
$GF(2^m)$ on FPGA using CRC and hamming codes.  IEEE
Trans. Reliability, to appear 2023.

\bibitem{key-80}A. Cintas-Canto, M. Mozaffari-Kermani, and R. Azarderakhsh. Reliable architectures for composite-field-oriented constructions
of McEliece post-quantum cryptography on FPGA.  IEEE Transactions
on Computer-Aided Design Integr. Circuits Syst., vol. 40, no. 5, pp.
999-1003, 2021.

\bibitem{key-81}A. Cintas-Canto, M. Mozaffari-Kermani, and R. Azarderakhsh. Reliable constructions for the key generator of code-based post-quantum
cryptosystems on FPGA.  ACM Emerging Technologies in Computing Systems
(special issue on CAD for Hardware Security), vol. 29, no. 1, pp.
5:1-5:20, 2023.

\bibitem{key-76}S. Huang, R. Sim, C. Chuengsatiansup, Q. Guo, T. Johansson. Cache-timing attack against HQC. ePrint, 2023.

\bibitem{key-77}T. Schamberger, J. Renner, G. Sigl, and A. Wachter-Zeh. A power side-channel attack on the CCA2-secure HQC KEM. CARDIS 2020, pp. 119-134, Springer, 2021.

\bibitem{key-78}T. Schamberger, L. Holzbaur, J. Renner, A. Wachter-Zeh,
and G. Sigl. A power side-channel attack on the reed-muller reed-solomon
version of the HQC cryptosystem. PQCrypto, pp. 327-352,
Springer, 2022.

\bibitem{key-72}G. Goy, A. Loiseau, and P. Gaborit. A new key
recovery side-channel attack on HQC with chosen ciphertext. PQCrypto 2022, pp. 353-371, Springer, 2022.
\end{thebibliography}
\end{document}